\documentclass{article}

\usepackage{arxiv}

\usepackage[utf8]{inputenc} 
\usepackage[T1]{fontenc}    
\usepackage{hyperref}       
\usepackage{url}            
\usepackage{booktabs}       
\usepackage{amsfonts}       
\usepackage{nicefrac}       
\usepackage{microtype}      
\usepackage{graphicx}

\usepackage{lmodern}
\usepackage[numbers]{natbib}
\usepackage{multirow}

\title{Enhanced Forest Inventories for Habitat Mapping: A Case Study in the Sierra Nevada Mountains of California}

\author{
    Maxime Turgeon\\
  \href{https://www.tesera.com/}{Tesera Systems Inc.}\\
  and\\
  University of Manitoba\\
   \And 
   Michael Kieser\\
  \href{https://www.tesera.com/}{Tesera Systems Inc.}\\
   \And
   Dwight Wolfe\\
  \href{https://www.tesera.com/}{Tesera Systems Inc.}\\
   \And
   Bruce MacArthur\\
  \href{https://www.tesera.com/}{Tesera Systems Inc.}\\
}

\begin{document}
\maketitle



\begin{abstract}
Traditional forest inventory systems, originally designed to quantify merchantable timber volume, often lack the spatial resolution and structural detail required for modern multi-resource ecosystem management. In this manuscript, we present an Enhanced Forest Inventory (EFI) and demonstrate its utility for high-resolution wildlife habitat mapping. The project area covers 270,000 acres of the Eldorado National Forest in California's Sierra Nevada. By integrating 118 ground-truth Forest Inventory and Analysis (FIA) plots with multi-modal remote sensing data (LiDAR, aerial photography, and Sentinel-2 satellite imagery), we developed predictive models for key forest attributes. Our methodology employed a two-tier segmentation approach, partitioning the landscape into approximately 575,000 reporting units with an average size of 0.5 acre to capture forest heterogeneity. We utilized an Elastic-Net Regression framework and automated feature selection to relate remote sensing metrics to ground-measured variables such as basal area, stems per acre, and canopy cover. These physical metrics were translated into functional habitat attributes to evaluate suitability for two focal species: the California Spotted Owl (\textit{Strix occidentalis occidentalis}) and the Pacific Fisher (\textit{Pekania pennanti}). Our analysis identified 25,630 acres of nesting and 26,622 acres of foraging habitat for the owl, and 25,636 acres of likely habitat for the fisher based on structural requirements like large-diameter trees and high canopy closure. The results demonstrate that EFIs provide a critical bridge between forestry and conservation ecology, offering forest managers a spatially explicit tool to monitor ecosystem health and manage vulnerable species in complex environments.
\end{abstract}


\section{Introduction}

Historically, forest inventory systems were primarily designed for the purpose of quantifying merchantable sawtimber volume. Past conventional sampling designs relied on sparse grids with low ground plot intensity to generate broad statistical aggregates at the planning unit or forest landscape level. While these data were used for regional biometric modeling, they lacked the spatial resolution necessary to inform forest managers about the spatial arrangement of the forest vegetation, tree species composition, and they provided limited forest structure attribute data~\citep{smith2002forest}. Because of the relative abundance of forests across North America, these broad statistical summaries were seen as sufficient and adequate for the purposes of managing timber forests.

Today's forests are managed for a broader array of social, economic, environmental and cultural values. Thanks to advancements in remote-sensing technologies and computational processing, Enhanced Forest Inventories (EFI) can provide a much finer granularity to estimates of forest structure and composition. By leveraging LiDAR (Light Detection and Ranging) and high-resolution multispectral imagery, coupled with field-calibrated biometric models, these systems generate a more comprehensive attribution and finer depiction of vegetation and forest structure~\citep{white2025enhanced}. This granularity allows for the interrogation of forest structure at a scale that is closer to that at which wildlife interact with their environment.


In this manuscript, we present an EFI over a 270,000-acre subset of the Eldorado National Forest, situated in the Central Sierra Nevada, approximately 50 miles east of Sacramento. The terrain of the project area is characteristic of the western Sierra slope, featuring a complex topography of river canyons, ridges, and mid-elevation plateaus~\citep{stephens2015historical}. The dominant vegetation type is the Sierra Mixed Conifer forest, a globally distinct ecosystem noted for its high biomass and species diversity~\citep{stephens2018historical}.

The main objective of this manuscript is two-fold: to describe how an EFI dataset can be generated from remote-sensing data; and to demonstrate how it can be used for habitat mapping. By translating physical metrics (e.g.\ trees per acre, diameter at breast height, and basal area) into functional habitat attributes, we can evaluate the capacity of the landscape to support targeted avian and mammalian guilds. This process serves as a potential bridge between forest inventory data and the desired forest habitat conditions articulated in forest management plans.

The remainder of this manuscript is structured as follows: in Section~\ref{sec:dataset}, we describe the process by which the remote-sensing datasets and FIA plots were converted into an EFI over the project area. In Section~\ref{sec:habitat}, we describe how the EFI can be used to map the probable habitat of the California Spotted Owl (\textit{Strix occidentalis occidentalis}) and the Pacific Fisher (\textit{Pekania pennanti}), two important yet vulnerable species in the Sierra Nevada ecosystem. Finally, we discuss the results in Section~\ref{sec:discussion}, highlighting potential improvements and future directions.

\section{Dataset}\label{sec:dataset}

\begin{figure}
 \centering
        \includegraphics[width=\textwidth]{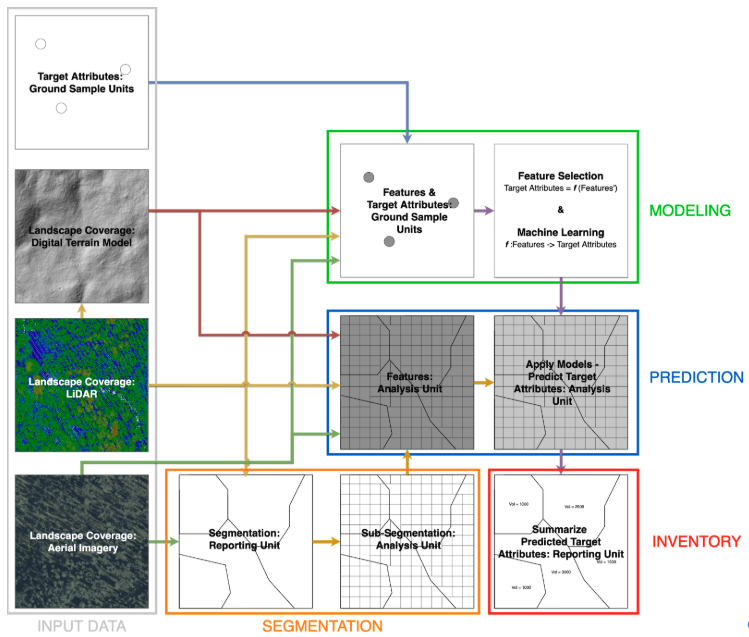}
 \caption{Visual representation of the EFI generation process.}
\label{figprocess}
\end{figure}

In this Section, we describe the process of generating an EFI using remote sensing datasets. We present the various datasets that were collected and processed, and we highlight the different computational steps required for creating the high-resolution forest inventory. A visual representation of the entire process appears in Figure~\ref{figprocess}. The final dataset is available at DOI: \url{https://doi.org/10.5281/zenodo.18020476} in the form of a GeoPackage file.

\subsection{Project Area}

This analysis focuses on an area of approximately 270,000 acres within the Eldorado National Forest (ENF), situated in the Central Sierra Nevada region of California. The project area is located approximately 50 miles east of Sacramento and covers nearly 50\% of the central-western portion of the broader 570,000-acre forest. The study area holds deep cultural and historical significance, lying within the ancestral territories of the Miwok, Washoe, and Nisenan people~\citep{enfwebsite}. It is also an important part of the habitat of the California Spotted Owl and the Pacific Fisher~\citep{gutierrez2017california}.

The ENF is characterized by its complex environment and mountainous topography~\citep{enfwebsite}. Bordered by the Tahoe National Forest to the north and the Stanislaus National Forest to the south, the landscape ranges in elevation from 1,000 feet in the western foothills to over 10,000 feet above sea level along the Sierra crest. This vertical diversity facilitates a broad range of vegetation types, including woodland, chaparral, mixed conifer, true fir, and subalpine forests. The physical geography is defined by the steep canyons of the Mokelumne, Cosumnes, American, and Rubicon rivers, which alternate with plateaus of moderate relief. Such topographic complexity creates distinct microclimates and moisture regimes, with the forest receiving an average of 56 inches of precipitation annually. Due to the heterogeneity of the forest stand structure and biomass distribution across the landscape, a high-resolution modeling approach is required in order to accurately model the forest and the various habitats.

\subsection{Remote Sensing and FIA Plots}

The foundation of the EFI framework is an integration of ground-truth observations and multi-modal remote sensing platforms~\citep{white2016remote}. In the context of this analysis, this corresponds to ground plot data, imagery, and LiDAR data. Ground plot data were obtained from the Forest Inventory and Analysis (FIA) program of the United States Department of Agriculture (USDA) Forest Service~\citep{burrill2024forest}. Within the project boundaries, 118 FIA plots were identified, and the data from the most recent collection cycle (from 2014 to 2019) was compiled for analysis. We utilized several core tables from the FIA DataMart, including the PLOT, PLOTGEOM, SUBPLOT, COND, and TREE tables, to establish a link between individual tree measurements and their precise geospatial locations. From this tree level information, we compiled several forest attributes, including average tree height, canopy cover, stem density, and average diameter at breast height; a complete list of attributes is presented in Table~\ref{tab:efi_attributes}.

\begin{table}[ht]
\centering
\caption{Forest Attributes Predicted by the EFI Model}
\label{tab:efi_attributes}
\begin{tabular}{lll}
\toprule
\textbf{Attribute} & \textbf{Description} & \textbf{Units} \\
\midrule
\texttt{pred\_bapa} & Basal area & Square feet per acre ($ft^2/ac$) \\
\texttt{pred\_bapa\_softwood} & Basal area covered by softwood species & Square feet per acre ($ft^2/ac$) \\
\texttt{pred\_bapa\_snag} & Basal area covered by snags & Square feet per acre ($ft^2/ac$) \\
\texttt{pred\_ht} & Average height & Feet ($ft$) \\
\texttt{pred\_dia} & Average diameter & Inches ($in$) \\
\texttt{pred\_tpa} & Number of trees per area & Stems per acre ($count/ac$) \\
\texttt{pred\_cagpa} & Above-ground carbon density & Tons per acre ($tons/ac$) \\
\texttt{pred\_cncvr\_pct} & Canopy cover & Percentage (\%) \\
\bottomrule
\end{tabular}
\end{table}

The remote sensing data employed for this study includes high-density LiDAR, aerial photography, and multi-spectral satellite imagery. LiDAR data were sourced from the USGS 3DEP (United States Geological Survey 3D Elevation Program). Data collection over the area was performed in 2019 under the USGS \texttt{CA\_UpperSouthAmerican\_Eldorado\_2019\_B19} project and meets the USGS QL1 quality level (i.e.\ vertical accuracy of 10 cm, nominal pulse spacing of at most 0.35 m, nominal pulse density of at least 8 pts/square meter). Complete processing of this dataset involved the processing of over 2,400 LAZ files totaling approximately 628.6 GB; we used the \texttt{lidR} and the \texttt{sf} R packages~\citep{lidr,sf}. To complement this structural data, we also processed 0.3m resolution four-band aerial imagery from the National Agriculture Imagery Program (NAIP) (resampled to 1m resolution), and 10m resolution 13-band Sentinel-2 L2A satellite imagery. These imagery datasets were both collected in the summer of 2020. The Sentinel-2 data is particularly critical for capturing the spectral signatures necessary for land cover classification, species identification, and the differentiation between live and dead biomass~\citep{phiri2020sentinel}. By combining the structural precision of LiDAR with the spectral depth of Sentinel-2, the model can more effectively distinguish between complex forest types and successional stages~\citep{lefsky2002lidar}.

\subsection{Segmentation}

A significant challenge in EFI development is the Modifiable Areal Unit Problem (MAUP), where the scale and shape of the spatial units can influence the results of the statistical analysis~\citep[Chapter 1.22]{huang2017comprehensive}~\citep{maupwebsite}. To mitigate this, our methodology employs a two-step segmentation approach. First, the landscape is partitioned into \emph{analysis units} that approximate the size and geometry of the physical ground plots. This ensures that the machine learning models are deployed over spatial units that are ecologically and statistically comparable to the FIA plots.

\begin{figure}
 \centering
        \includegraphics[width=\textwidth]{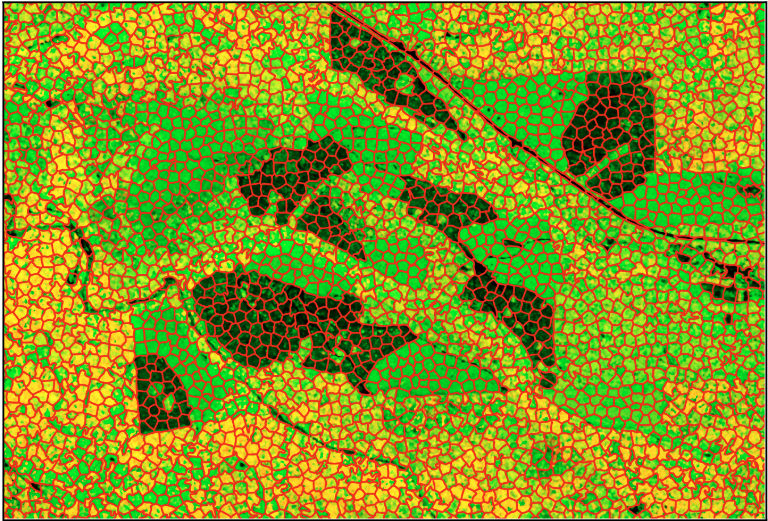}
 \caption{Segmentation of the project area into 0.5-acre polygons.}
\label{figsegmentation}
\end{figure}

The second tier involves the creation of \emph{reporting units}, which are derived from a combination of the LiDAR-based Digital Terrain Model (DTM), the canopy surface, and NAIP-derived vegetation indices (e.g. Normalized Difference Vegetation Index). An automated segmentation process was utilized to divide the 270,000-acre project area into approximately 575,000 unique polygons. These polygons average 0.5 acre in size and are designed to capture forest patches of uniform canopy height and vegetation type. See Figure~\ref{figsegmentation} for an illustration of these polygons.

\subsection{Modelling and Prediction}

The next step is to relate the remote-sensing data with the ground-plot data using machine learning. First, we summarized the imagery and LiDAR datasets over the analysis units by generating hundreds of summary statistics that represent localized forest conditions. These features include various summaries of the LiDAR point densities at multiple height strata, topographic indices derived from the DTM, and spectral vegetation indices (e.g.\ NDVI, EVI) and band ratios calculated from both NAIP and Sentinel-2 data~\citep{yan2025global}. These features were then normalized to ensure that they are all on a comparable scale.

To reduce the ``curse of dimensionality'', we applied statistical feature selection methods to prune the initial pool of approximately 1,600 features down to the 700 most informative variables. This reduction minimizes computational overhead and limits the risk of model overfitting, ensuring that only those features with a strong relationship to the target forest attributes are retained~\citep{guyon2003introduction}.

Next, for the predictive modeling phase, we employed an Elastic-Net Regression framework to relate the compiled ground plot attributes to the normalized remote sensing features. Elastic-net is particularly effective for forest inventory applications because it combines the penalties of ridge and lasso regression, allowing for effective variable selection even in the presence of highly correlated features~\citep{zou2005regularization}. To ensure the robustness of our results, we utilized a comprehensive grid search combined with cross-validation to select the optimal tuning parameters for each model~\citep[Chapter 7]{hastie2009elements}. Model performance was evaluated using metrics such as the Root Mean Square Error (RMSE) and the coefficient of determination ($R^2$). The final models demonstrated strong predictive capabilities across all targeted forest attributes.

\begin{figure}
 \centering
        \includegraphics[width=\textwidth]{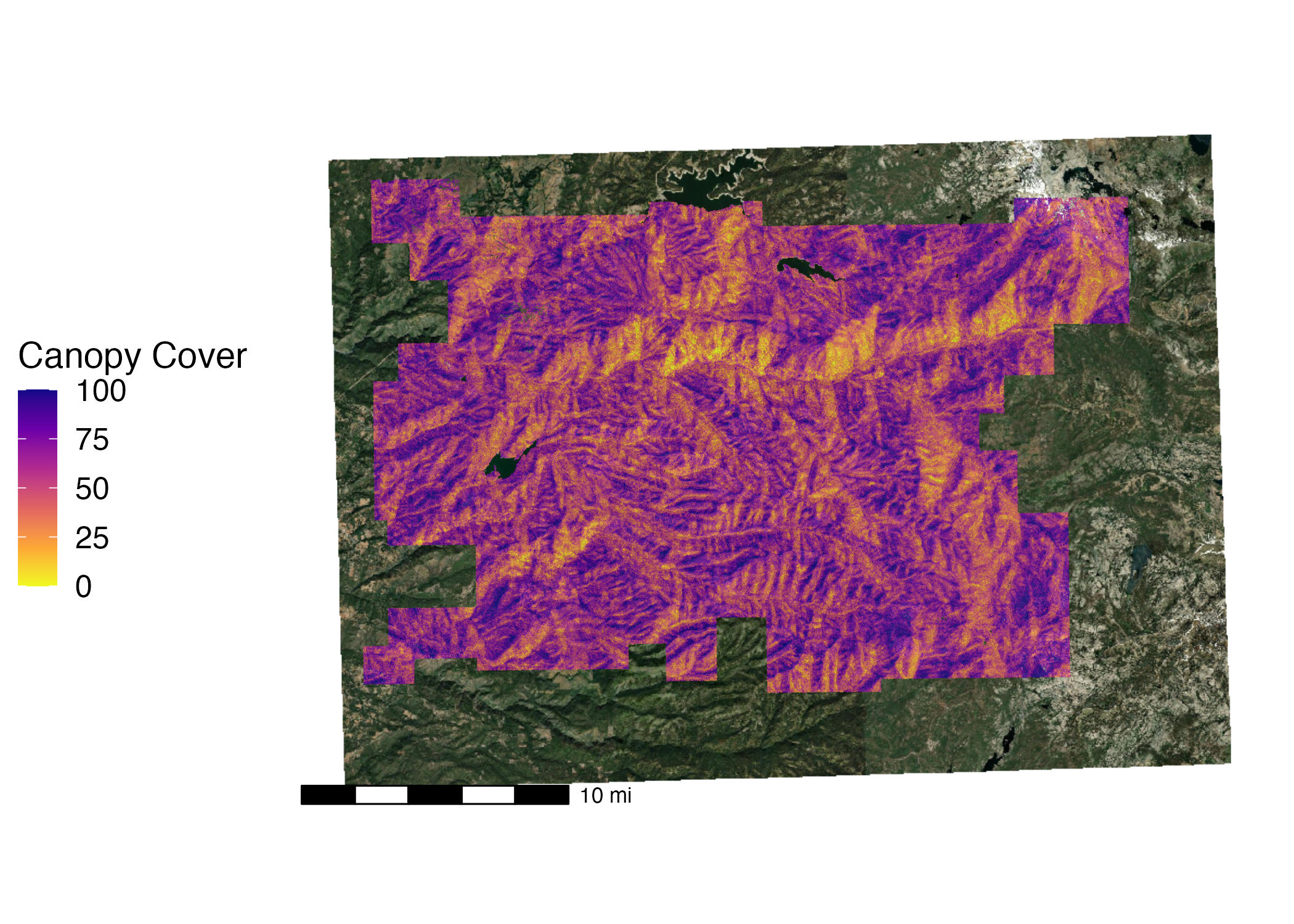}
 \caption{Canopy Cover (\%) over the project area.}
\label{figcanopy}
\end{figure}

\begin{figure}
 \centering
        \includegraphics[width=\textwidth]{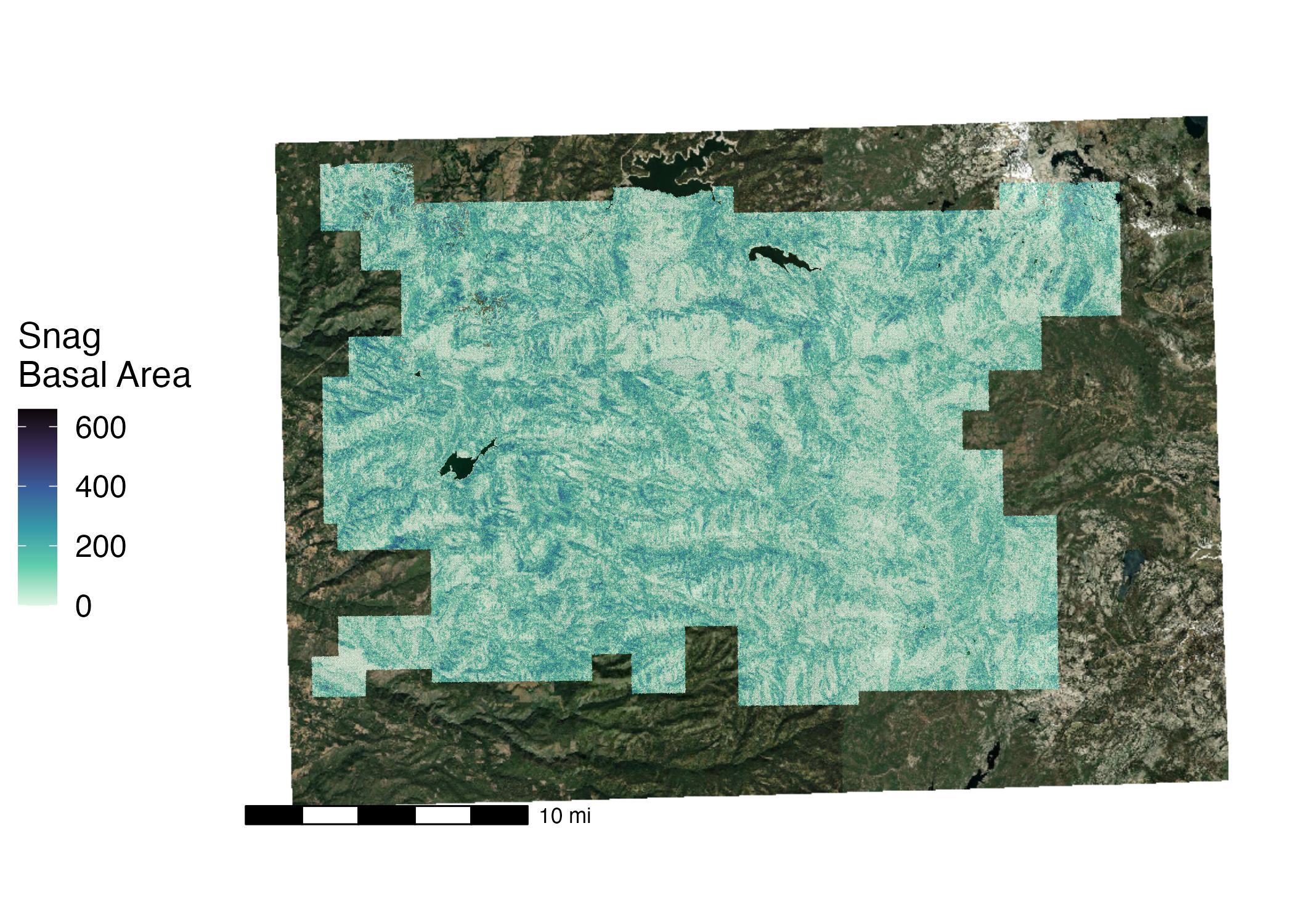}
 \caption{Basal area covered by snags (square feet/acre) over the project area.}
\label{figsnag}
\end{figure}

Once the models were finalized, they were used to predict the forest attributes over every single analysis unit. Predictions from analysis units were further summarized over the corresponding reporting units using an area-weighted average, thus providing a spatially continuous inventory of the Eldorado National Forest. Figures~\ref{figcanopy} and \ref{figsnag} show the spatial distribution of two of these attributes over the project area.

\section{Habitat Mapping}\label{sec:habitat}

The transition toward high-resolution precision forestry represents a fundamental shift in how forest ecosystems are monitored and managed. As described above, this methodology provides granular, polygon-level data at a resolution of approximately 0.5 acre, capturing key quantitative attributes of forest structure including Trees Per Acre (TPA), Diameter at Breast Height (DBH), and Canopy Cover (CC). This spatial understanding is critical for assessing the ecosystem health and structural capacity of the 270,000-acre project area to support high-quality habitat for targeted avian and mammalian fauna of the Central Sierra Nevada.

The utility of an EFI for forest management has already been documented in the literature~\citep{white2016remote,white2025enhanced}. The goal of this analysis is to demonstrate how an EFI can also be used for habitat mapping. Specifically, this analysis focuses on two primary guilds sensitive to structural changes in the forest: the Late-Seral/Old Forest Guild (Birds) and the Forest Carnivore Guild (Mammals). For demonstration purposes, the Late-Seral guild will be represented by the California Spotted Owl, whose presence is strongly associated with complex, multi-layered old-growth coniferous forests~\citep{gutierrez2017california}. The Forest Carnivore guild will be represented by the Pacific Fisher, a species that depends on high-quality structural elements for thermal regulation, resting, and denning. The ability of the EFI to quantify current forest structure against the specific requirements of these focal species allows for a direct evaluation of habitat quality and spatial distribution. 

\subsection{The Late-Seral/Old Forest Guild: California Spotted Owl}

The California Spotted Owl is a habitat specialist requiring older, closed-canopy forests with complex vertical structures~\citep{verner1992california}. High-suitability nesting and roosting habitat for the California Spotted Owl is defined by strict quantitative thresholds. Functional nesting habitat requires a total canopy cover of at least 60\%, with the dominant and co-dominant canopy closure reaching at least 40\%~\citep{keane2014california,tempel2016meta}. Furthermore, polygons must contain an average density of greater than nine stems per acre of conifers with a DBH greater than 35 inches. These large trees, often featuring broken tops or heartwood decay, provide the necessary platforms for nesting and roosting~\citep{blakesley2005site}. While nesting sites are rigid in these requirements, foraging habitat shows more flexibility, potentially occurring in areas with 40\% canopy cover or increased structural heterogeneity.

In our dataset, we have TPA, DBH, and CC; however, they represent an overall average and therefore they do not meet the exact requirements above. To operationalize the habitat requirements above, we designated as ``Nesting'' the polygons for which \texttt{pred\_tpa} was larger than 9, \texttt{pred\_cncvr\_pct} was larger than 60, \texttt{pred\_bapa\_softwood} was larger than half of \texttt{pred\_bapa} (i.e.\ polygons where at least half the basal area is accounted for by softwood species), and \texttt{pred\_dia} larger than 25. This last criterion was chosen as it represents the third quartile of the average DBH for the entire project area; moreover, polygons with a larger value of \texttt{pred\_dia} are more likely to contain trees with large diameters. Finally, we designated as ``Foraging'' the polygons which met the criteria above, except that we relaxed \texttt{pred\_cncvr\_pct} to be larger than 40. Every other polygon was designated as ``Unlikely''.

\begin{figure}
 \centering
        \includegraphics[width=\textwidth]{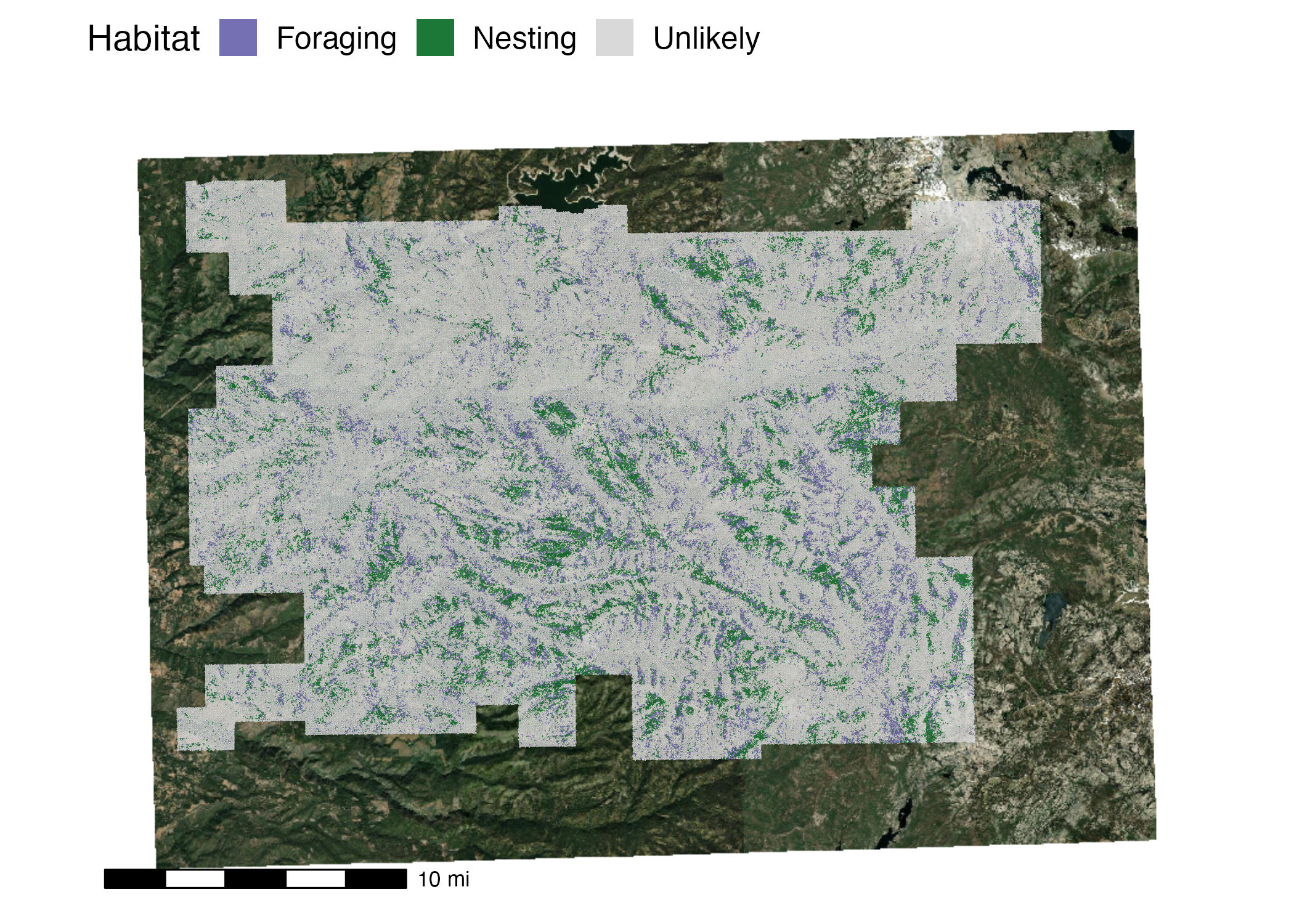}
 \caption{Habitat suitability for the California Spotted Owl over the project area.}
\label{figcso}
\end{figure}

Figure~\ref{figcso} shows a map of the probable habitat for the California Spotted Owl. The foraging polygons account for 26,622 acres, and the nesting polygons account for 25,630 acres.

\subsection{The Forest Carnivore Guild: Pacific Fisher}

Habitat suitability for the Pacific Fisher is primarily limited by the availability of resting and denning structures. The Pacific Fisher, at the southern extent of its range in the Sierra Nevada, relies heavily on canopy cover for thermal buffering against temperature extremes~\citep{davis2007regional}. High-suitability habitat requires a minimum canopy cover of approximately 60\%. Resting sites typically occur in cavities or platforms within live trees or snags that are among the largest available, often exceeding 35 inches DBH~\citep{zielinski2012estimating}. Fishers demonstrate a specific preference for large, shade-intolerant species such as Ponderosa and Sugar Pines~\citep{purcell2009resting}.

To operationalize the habitat requirements above, we followed a similar process as for the California Spotted Owl: we designated as ``Likely'' the polygons for which \texttt{pred\_cncvr\_pct} was larger than 60, \texttt{pred\_bapa\_softwood} was larger than half of \texttt{pred\_bapa}, and \texttt{pred\_dia} larger than 25. Every other polygon was designated as ``Unlikely''.

\begin{figure}
 \centering
        \includegraphics[width=\textwidth]{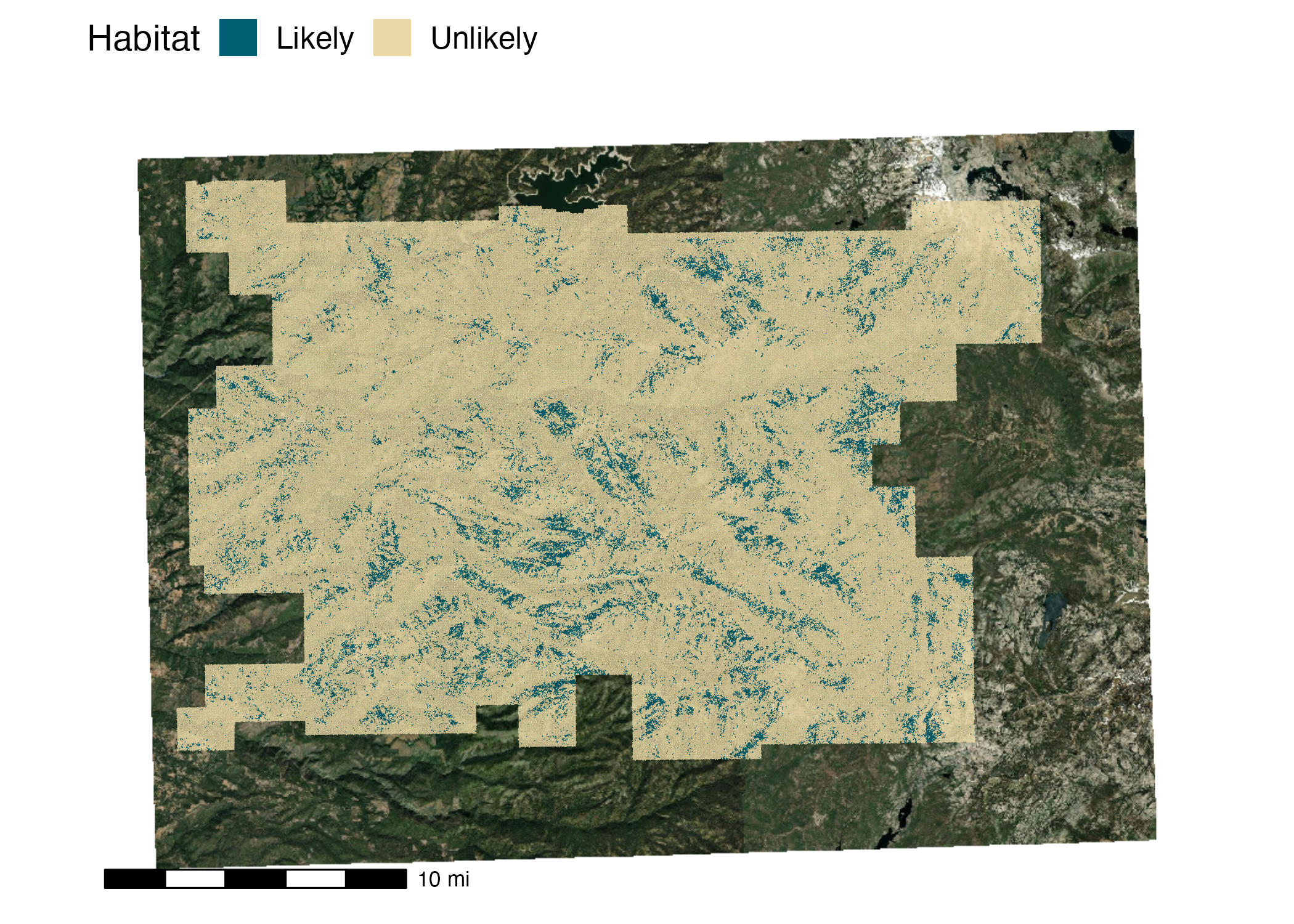}
 \caption{Habitat suitability for the Pacific Fisher over the project area.}
\label{figpf}
\end{figure}

Figure~\ref{figpf} shows a map of the probable habitat for the Pacific Fisher. The likely habitat polygons account for 25,636 acres. Table~\ref{tab:habitat_criteria} summarizes the criteria for both guilds.

\begin{table}[ht]
\centering
\caption{Habitat Suitability Criteria for Target Species}
\label{tab:habitat_criteria}
\begin{tabular}{llcccc}
\toprule
\textbf{Species} & \textbf{Habitat Type} & \texttt{pred\_cncvr\_pct} & \texttt{pred\_tpa} & \texttt{pred\_dia} & \textbf{Softwood \%}* \\
\midrule
\multirow{2}{*}{California Spotted Owl} & Nesting & $> 60\%$ & $> 9$ & $> 25$ in & $> 50\%$ \\
                             & Foraging & $> 40\%$ & $> 9$ & $> 25$ in & $> 50\%$ \\
\midrule
Pacific Fisher               & Likely   & $> 60\%$ &       & $> 25$ in & $> 50\%$ \\
\bottomrule
\multicolumn{6}{l}{\small *Percentage of total basal area accounted for by softwood species.}
\end{tabular}
\end{table}

\section{Discussion}\label{sec:discussion}

This manuscript demonstrates how an EFI can be used for habitat mapping. The results of this analysis show that the transition toward high-resolution precision forestry, facilitated by EFIs, provides a fundamental shift in how forest ecosystems are monitored and managed. By leveraging LiDAR and high-resolution multispectral imagery, we generated a granular, polygon-level dataset at a half-acre resolution for 270,000 acres of the Eldorado National Forest. This precision allows for a spatial understanding that is critical for assessing the ecosystem health and structural capacity of the landscape to support targeted avian and mammalian fauna. The primary utility of this EFI approach is the ability to translate physical metrics, such as Trees Per Acre (TPA), Diameter at Breast Height (DBH), and Canopy Cover (CC), into functional habitat attributes. 

Our mapping focused on two primary guilds sensitive to structural changes: the Late-Seral/Old Forest Guild and the Forest Carnivore Guild. The California Spotted Owl acts as an indicator for complex, multi-layered old-growth coniferous forests. High-suitability nesting habitat requires strict thresholds. The Pacific Fisher relies on high-quality structural elements for thermal regulation, resting, and denning. Suitability for this species is primarily limited by the availability of large structures and a minimum canopy cover of approximately 60\% to provide thermal buffering against temperature extremes. The ability of the EFI to quantify current forest structure against these specific requirements allows for a direct, spatially explicit evaluation of habitat quality.

As of late 2025, the conservation status of these focal species underscores the importance of accurate habitat mapping. The California Spotted Owl is currently proposed for federal listing under the Endangered Species Act (ESA), with the Sierra Nevada population proposed as \textit{Threatened}. Similarly, the Pacific Fisher faces a bifurcated regulatory landscape: the Southern Sierra Nevada population is federally and state-listed as \textit{Endangered} while the Northern California population is state-listed as \textit{Threatened} but lacks federal listing as of late 2025. The high-resolution inventory presented here serves as a bridge between current forest conditions and the desired forest conditions articulated in management plans, providing a tool to manage these vulnerable species effectively.

\subsection{Limitations and Future Directions}

Despite the high level of detail provided by this EFI, several limitations remain. Data collection for this study was spread over multiple seasons; ideally, all remote sensing and ground plot data would be synchronized to reduce measurement error. Furthermore, while we utilized 118 FIA plots, an increased number of plots would have enabled more robust modeling of specific species composition (e.g.\ basal area covered by Douglas Fir or Ponderosa Pine). Current LiDAR pulse densities allowed for accurate canopy modeling, but higher densities would be required to penetrate further below the canopy to model basal area of snags with minimum diameters and coarse woody debris, the latter of which was omitted from this analysis due to modeling challenges. 

Future iterations of this work could transform the static EFI into a probabilistic habitat model by integrating additional training data (e.g.\ documented presence of California Spotted Owls or Pacific Fishers). Such advancements would further refine the capacity of forest managers to understand the physical environment that sensitive wildlife use for shelter, foraging, and breeding. Finally, the exclusion of human footprint data remains a constraint on the comprehensiveness of habitat mapping; a future iteration could integrate this information into the analysis above~\citep{burdett2010interfacing}.

\section{Conclusion}

This manuscript introduces an EFI over a large portion of the Eldorado National Forest in California. We explained how it can be produced by combining remote-sensing data, ground plot measurements, and machine learning. We then showed how this level of detail can be used to assess habitat suitability for two important species in the Sierra Nevada ecosystem: the California Spotted Owl and the Pacific Fisher.

Through this work, our objective was to provide a bridge between the disciplines of forestry and ecology. As forest managers continue to adopt the EFI methodology across North America, we believe there is an opportunity for including local conservation groups within their planning discussions. We hope that this dataset will facilitate future collaborations between forest managers and ecologists, ultimately leading to better-informed decisions for both forest management and biodiversity conservation.


%
%




\section*{Data}
The EFI dataset is available at DOI: \url{https://doi.org/10.5281/zenodo.18020476}. It is released under the Creative Commons CC-BY 4.0 license. Additionally, the R code for processing data and generating the figures is released under the MIT License and is hosted at \url{https://github.com/tesera/hris-habitat-mapping}, which also includes a direct link to the dataset.



\bibliographystyle{unsrtnat}
\bibliography{references}

\end{document}